\documentclass{aastex} 
\usepackage{emulateapj5}
\usepackage{apjfonts}

\newcommand{\kms}{\ensuremath{\mathrm{km}\;\mathrm{s}^{-1}}}

\newcommand{\oii}{[\ion{O}{2}]}
\newcommand{\oiii}{[\ion{O}{3}]}
\newcommand{\ovii}{\ion{O}{7}}
\newcommand{\oviii}{\ion{O}{8}}

\newcommand{\neix}{\ion{Ne}{9}}
\newcommand{\nex}{\ion{Ne}{10}}

\newcommand{\halpha}{H\ensuremath{\alpha}}
\newcommand{\hbeta}{H\ensuremath{\beta}}

\newcommand{\oursnr}{1E\kern0.4em 0102.2-7219}

\newcommand{\mysim}{\ensuremath{\sim\kern-0.3em}}

\newcounter{hours}\newcounter{minutes}
\newcommand{\printtime}{%
  \setcounter{hours}{\time/60}%
  \setcounter{minutes}{\time-\value{hours}*60}%
  \thehours : \kern-0.4em \theminutes}

\newcommand\plotonefiddle[2]{%
 \centering
 \leavevmode
 \includegraphics[angle=#2,width={\columnwidth}]{#1}%
}%

\shortauthors{Gaetz et al.}
\shorttitle{Arcsecond X-ray Imaging of \oursnr}

\begin{document}

\submitted{Received 2000 January 21; accepted 2000 March 13}

\title{{\it Chandra X-ray Observatory\/} Arcsecond Imaging of the Young, Oxygen-rich Supernova Remnant \oursnr}


\author{
T. J. Gaetz, Yousaf M. Butt, Richard J. Edgar, Kristoffer A. Eriksen, \\
Paul P. Plucinsky, Eric M. Schlegel, and Randall K. Smith
}
\affil{Harvard-Smithsonian Center for Astrophysics, 60 Garden Street, Cambridge, MA 02138}

\begin{abstract}
We present observations of the young, oxygen-rich supernova remnant 
\oursnr\ taken by
the {\it Chandra X-ray Observatory\/} during its orbital 
activation and checkout
phase.  The boundary of the blast-wave shock is clearly seen for the 
first time, allowing the diameter of the remnant and 
the mean blast-wave velocity to be determined accurately.  
The prominent X-ray bright ring of material may be the result of 
the reverse shock 
encountering ejecta; the radial variation of \ovii\ versus \oviii\
emission indicates an ionizing shock propagating inward, possibly
through a strong density gradient in the ejecta.  We compare the
X-ray emission to Australia Telescope Compact Array 6 cm radio observations
(Amy and Ball 1993) and to archival
{\it Hubble Space Telescope\/} {\oiii} observations.  The ring
of radio emission is predominantly inward of the outer blast-wave, which is\break
consistent with an interpretation of synchrotron radiation originating 
behind the blast-wave but outward of the
bright X-ray ring of emission.  Many (but not all) of the
prominent optical filaments are seen to correspond to X-ray bright
regions.  We obtain an upper limit of 
$\mysim 9\times 10^{33}\; \mathrm{erg}\; \mathrm{s}^{-1}$ ($3\sigma$) on any
potential pulsar X-ray emission from the central region.

\end{abstract}

\keywords{Magellanic Clouds --- shock waves --- supernova remnants --- X-rays: ISM }

\section{Introduction}
   \label{sec:introduction}

The supernova remnant \oursnr\ (hereafter E0102) in the 
Small Magellanic Cloud (SMC) is an
interesting example of a young O-rich supernova remnant.  It was 
first identified as a likely supernova remnant by \citet{sm81} based on
{\it Einstein\/} soft X-ray images of the SMC.
\citet{dtm81} found a filamentary \oiii\ shell ($\mysim 24\arcsec$ diameter);
the remnant was not visible in \halpha\ 
(\oiii/\hbeta\ $\gtrsim\kern-0.3em 60$), 
and they classified the remnant as O-rich.  
They noted diffuse \oiii\ and \halpha\ emission surrounding 
the \oiii\ filaments, with an emission ``hole'' of diameter 
$\mysim 35$--$40\arcsec$ surrounding the \oiii, and suggested that the diameter
of the hole could correspond to that of the blast-wave.
In a follow-up spectroscopic study \citep{td83}, O-rich material
was found to have a full width velocity dispersion of $\mysim 6500\;\kms$.
If the O-rich material is assumed to be undecelerated ejecta, 
the 6.9 \break
pc diameter of the \oiii\ emission (assuming a distance \break
of 59 kpc)
implies an age of $\mysim 1000$ yr.  More recently, the \break
remnant
has been studied with the {\it Hubble Space Telescope\/} \citep{b2000}.
The remnant has also been observed in the UV \citep{brdm89},
radio \citep{ab93} and numerous times in the X-ray 
(Hayashi et al. 1994 and references therein).
The X-ray spectrum of the remnant has proved difficult to understand;
\citet{hkmmth94} were unable to obtain a formally acceptable fit to the
{\it ASCA\/} data, even with a rather complicated nonequilibrium 
ionization model, and concluded that the abundances in the 
plasma are inhomogeneous.  By providing high spatial resolution, 
the {\it Chandra X-ray Observatory\/} will help us
to solve these problems by reducing the spatial confusion that 
hampered earlier X-ray studies.

\section{X-ray Data}
   \label{sec:x.ray.data}

\newcommand{\figone}{1}

We report here on observations of the supernova remnant E0102
obtained during the orbital activation and checkout 
period for the {\it Chandra X-ray Observatory\/} \citep{wos96}.  The data were
obtained with the S3 chip  of the Advanced CCD Imaging Spectrometer 
\citep[ACIS;][]{g92,b98},
a backside-illuminated CCD with low to moderate spectral resolution
($E/\Delta E \mysim 4.3$ at 0.5 keV, $\mysim 31$ at 5.9 keV).  
We analyzed observation Ids (ObsIds) 138 and 1231 (1999 Aug 23;
processing version R4C3UPD2) and ObsId 1423 
(1999 Nov 01; processing version R4CU4UPD1.1).
The 138 and 1231 data were taken with a focal plane 
temperature of $-100^\circ\mathrm{C}$; the 1423 data were obtained at
$-110^\circ\mathrm{C}$.  The effective exposure times were 9759,
9762 and 19104 s (ObsIds 138, 1231, and 1423, respectively).
We worked from {\it Chandra X-ray Observatory\/} Center (CXC) 
level 1 and level 2 event lists, and used the {\it ASCA\/}-like 
$[0,2,3,4,6]$ grade set.
The absolute sky coordinates derived from the FITS world 
coordinate system (E. W. Greisen \& M. Calabretta 2000, in preparation) data 
in the event list headers yielded slightly different positions for the 
three data sets; ObsIds 138 and 1231
gave consistent results, but ObsId 1423 is shifted 
$\mysim 1\farcs 6$ south-southeast relative to ObsId 138,
consistent with a recently discovered secular drift in {\it Chandra\/}
coordinate determinations between 1999 August and November.
The 138 and 1231 data were taken within about a day of
a boresight calibration observation and should be more accurate.
We registered the three data sets against each other and
applied the ObsId 138 coordinate transform.
The event data were binned on $0\farcs 492$ pixels (ACIS pixels) and 
smoothed with a $0\farcs 492$ FWHM Gaussian 
(Fig. \figone).
We computed radial profiles based on $20\arcdeg$ sectors and
$0\farcs 25$ radial binning centered on 
$\alpha = 1^\mathrm{h}\; 04^\mathrm{m}\; 1\fs 94$,
$\delta = -72\arcdeg 01\arcmin 52\farcs 0$ (J2000), the
X-ray centroid.

The X-ray emission is dominated by a bright ring with a \break 
radius of
$\mysim\kern-0.05em 14\arcsec$.  The {\it Einstein\/} and {\it ROSAT\/} HRIs
also detect \break
evidence for a ring of emission, but with at
least 5 times poorer spatial resolution,
the remarkable substructure and brightness variations were not as evident 
as with {\it Chandra}.  
The ring is inhomogeneous and is brightest and spatially sharpest 
(\mbox{$\lesssim\kern-0.3em 2\arcsec$} FWHM) in the southeast, 
with significant knots south-southwest
and west and a ridge northwest; the peak brightness varies by
a factor of about 3 around the ring.  The ring FWHM varies from 
$\mysim 2\arcsec$ to $\mysim 8 \arcsec$ with mean $\mysim 5\arcsec$
(approximately half the total counts lie within the FWHM).
The ring falls off most steeply toward the center, and more gradually
outward with suggestions of irregularities or scalloping.
 
The remnant is bounded by a faint plateau with a relatively sharp
outer edge, particularly in the northeast.  This plateau shows a hint of 
limb brightening in the east and northeast;
the southwest boundary is more diffuse and 
irregular.  
Interior to the bright ring the remnant is faint and patchy
with suggestions of brighter ridges; a bright feature extends from near 
the center toward the bright region south-southwest.

\bigskip
\begin{minipage}[l]{0.46\textwidth}
  \plotonefiddle{total.ps}{270}
  \medskip
  \noindent\begin{minipage}{0.99\linewidth} 
    \vskip \abovecaptionskip\footnotesize\noindent
    Fig. 1. --- Grayscale image of the smoothed {\it Chandra\/} X-ray data
     (ObsIds 1423, 138, 1231); the square root of the intensity (in
     units of 
     $\mathrm{counts}\; \mathrm{arcsec}^{-2}\; \mathrm{s}^{-1}$) is plotted.
   \par\vskip \belowcaptionskip
  \end{minipage}
\end{minipage}

\newcommand{\figtwo}{2}
\newcommand{\figthree}{3}
\newcommand{\figfour}{4}

The backside-illuminated S3 chip has good low-energy detection 
efficiency,
but the energy response is nonlinear and tempera\-ture-dependent, particularly
below $\mysim 1 \;\mathrm{keV}$; in view of the uncertainties in the
response matrices at this stage of the cali\-bration, we defer
a detailed spectral investigation and \break
abundance determination.
Nevertheless, the energy scales
are sufficiently accurate to allow identification of the major
spectral
features seen in the earlier {\it ASCA\/} data, and for qualitative
assessments to be made.  In
Figure~\figtwo\ ,
we plot the pulse-height amplitude (PHA) spectrum for the
remnant as a whole (combined ObsIds 138 and 1231).

We have examined a number of spectra extracted
from various regions in the remnant including the outer shock region, regions of
the bright ring, and brighter and fainter regions interior to the bright ring.
The spectra vary throughout the remnant.  The O and Ne lines
are less prominent in the outer shock region; the O and Ne seem to be
in a higher excitation state with much of the emission coming from the
H-like stages.  The plasma state (e.g., the ratio of the He-like 
to H-like O [and Ne]) also varies around the bright ring.

In order to get a better idea of the variation of plasma conditions
within the remnant, we constructed difference maps
based on bands centered around the prominent emission lines:
$\mysim 0.46$ -- $0.61$ keV (\ion{O}{7} He-like triplet),
$\mysim 0.61$ -- $0.72$ keV (\ion{O}{8} Ly$\alpha$),
$\mysim 0.86$ -- $0.98$ keV (\ion{Ne}{9} He-like triplet),
and $\mysim 0.98$ -- $1.10$ keV (\ion{Ne}{10} Ly$\alpha$).  In
Figure \figthree\,
we plot a difference image of the bands dominated by the
\oviii\ Ly $\alpha$ line and the \ovii\ He-like triplet; in
Figure~\figfour\
we show radial profiles
for a $20\arcdeg$ sector centered $40\arcdeg$ south of east 
(i.e., position angle $130\arcdeg$).

\citet{h88,h94} ruled out a simple uniform-density shell model for the
X-ray emission on the basis of the (much lower resolution) {\it Einstein\/}
and {\it ROSAT\/} HRI data.  The {\it Chandra\/} data also suggest 
that the X-ray morphology is more complex than a uniform shell.  In the top
panel of
Figure~\figfour, 
we plot the azimuthally averaged X-ray profile
and a best fit uniform shell model.
Overall, the {\it Chandra\/} data suggest a faint shell of X-ray emission
surrounding a bright inhomogeneous ring of emission, possibly thin 
and ribbon-like.  The X-ray remnant as a whole is remarkably round 
($\mysim 10$\% longer northeast to southwest than northwest to southeast).
The mean X-ray diameter is $\mysim 44\arcsec$.  At a distance 
of 59 kpc, this corresponds to a linear diameter of $\mysim 12.6$ pc;
for an assumed remnant age of $\mysim 1000$ yr, the mean blast-wave velocity
is $\mysim 6200\;\kms$.

\medskip
\begin{minipage}[l]{0.46\textwidth}
    \plotone{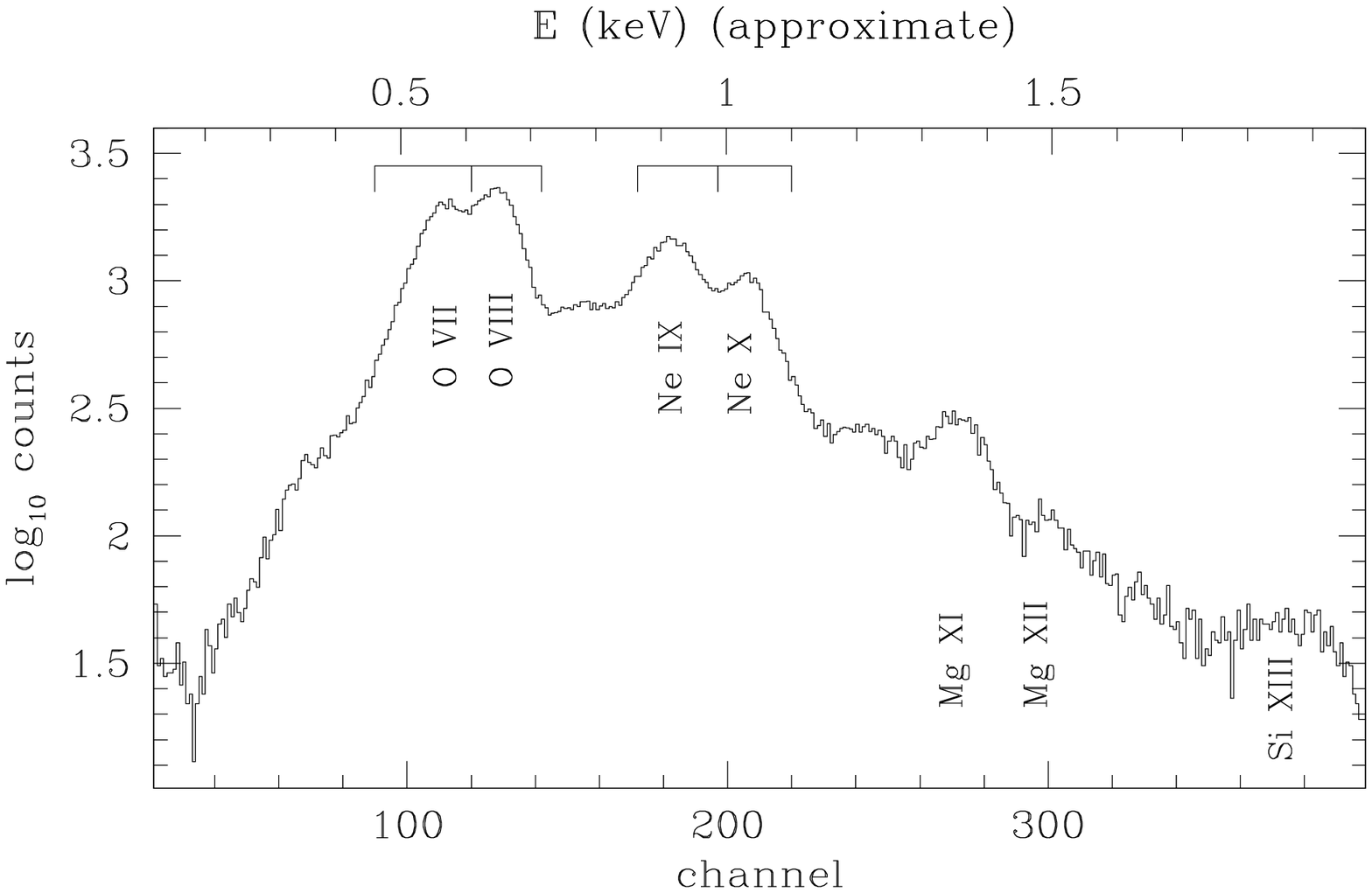}
    \medskip
    \noindent\begin{minipage}{0.99\linewidth} 
      \vskip \abovecaptionskip\footnotesize\noindent
     Fig. 2. ---
     {\it Chandra\/} low energy PHA spectrum for the whole remnant
     (ObsIds 138 + 1231).
     The upper axis is labeled with an {\it approximate\/} energy
     scale.  We also show the locations of the \ovii\-, \oviii\-,
     \neix\-, and \nex-dominated bands
     (for $T_\mathit{fp} = -100^\circ\;\mathrm{C}$;
     see text).  The approximate locations of the Ly$\alpha$ H-like
     and He-like triplet lines of O, Ne, Mg, and Si are indicated.
     \par\vskip \belowcaptionskip
    \end{minipage}
\end{minipage}
%

\medskip
\begin{minipage}[l]{0.46\textwidth}
  \plotonefiddle{dif_img.ps}{270}
  \medskip
  \noindent\begin{minipage}{0.99\linewidth} 
    \vskip \abovecaptionskip\footnotesize\noindent
   Fig. 3. ---
     Difference image of bands centered on the
     \oviii\ Ly $\alpha$ line and the \ovii\ He-like triplet.
     \oviii\ is white, and \ovii\ is black
     (in units of $\mathrm{counts}\; \mathrm{arcsec}^{-2}\; \mathrm{s}^{-1}$)
   \par\vskip \belowcaptionskip
  \end{minipage}
\end{minipage}

\medskip
\begin{minipage}[l]{0.46\textwidth}
  \plotone{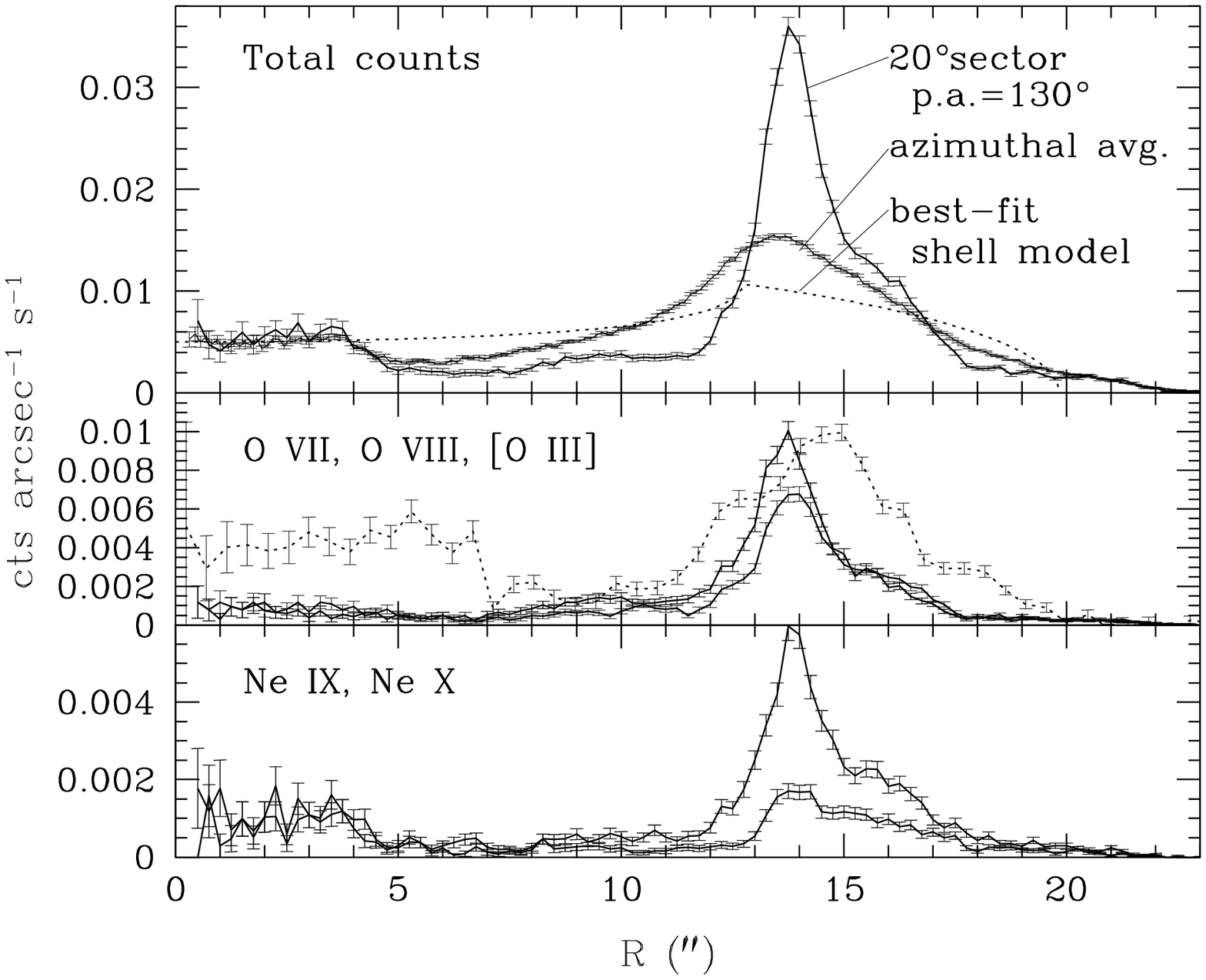}
  \medskip
  \noindent\begin{minipage}{0.99\linewidth} 
    \vskip \abovecaptionskip\footnotesize\noindent
   Fig. 4. ---
    Radial profiles for a 20$\arcdeg$-wide sector centered
    at position angle 130 (40$\arcdeg$ south of east; the narrowest
    and brightest portion of the X-ray ring).
    {\it Top\/}: Total count rate for the 20$\arcdeg$ sector.
    For reference, an azimuthal average over the remnant is shown,
    together with the best fit ($\chi^2/\mathrm{dof} = 230$) uniform
    shell model for the azimuthal average.
    {\it Middle\/}: \ovii\ ({\it upper curve\/}), \oviii\ ({\it lower curve\/}),
                  \oiii\ flux (arbitrary scale, {\it dotted curve\/})
                   (see \S\ref{sec:comparison.with.optical.data}.
    {\it Bottom\/}: \neix\ ({\it upper curve\/}), \nex\ ({\it lower curve\/}).
   \par\vskip \belowcaptionskip
  \end{minipage}
\end{minipage}
%

\section{Comparison with 6 cm Radio Observations}
   \label{sec:comparison.with.6.cm.radio.observations}

\newcommand{\figfive}{5}

\citet{ab93} observed the remnant with 3$\arcsec$ resolution at 6 cm using
the Australia Telescope Compact Array (ATCA), and found an outer shell of
radio emission with diameter $40\arcsec \pm 5\arcsec$.  They note that this
matches the size of the \oiii\ and \halpha\ ``emission hole'' of 
Dopita et al. (1981)
and the X-ray emission ({\it Einstein\/} HRI).  In 
Figure~\figfive,
we plot radio contours on the {\it Chandra\/}
X-ray image.  The radio ring maps out the outer X-ray emission well,
including the slight northeast to southwest elongation; the radio ring 
lies predominantly
outside the bright X-ray ring but within the outer X-ray rim.  The radio 
emission is brightest in
the northeast, just inside the brightest portion of the outer X-ray limb.
The radio ring weakens toward the southeast, where the X-ray ring is 
sharpest and brightest.  However, in the northwest, a bright radio ridge wraps 
somewhat around the brightest portion of an X-ray ridge and a region
of enhanced {\oiii} emission (see 
also \S\ref{sec:comparison.with.optical.data}).
In the southwest, a local enhancement of
the radio emission corresponds to a local X-ray enhancement, while 
the optical data indicate a corresponding gap in the 
\oii\ and \oiii\ emission.

\cite{ab93} noted a central enhancement in the 6 cm data and suggested the
possibility of a plerionic component (acknowledging the possibility
that it was a mere projection effect).  At the level of the available
{\it Chandra\/} aspect solution, the central radio enhancement overlaps a central
X-ray bright knot.  In the X-ray data, this knot coincides
with part of a bright ridge extending to the south (faintly visible in the
radio data as well), making it likely that the enhancement is
a chance superposition.
We obtain an X-ray upper limit on any plerionic contribution as follows:
After background subtraction, the number of counts above $\mysim 3$ keV
and within $10\arcsec$ of the center of the remnant is $30\pm 11$ counts
in a 9762 s exposure, or 
$\mysim (3.1 \pm 1.1) \times 10^{-3} \; \mathrm{counts} \;\mathrm{s}^{-1}$.  
Using an adopted Crab spectrum (with a power-law of index 2.05), the prelaunch
response matrix for the S3 chip, and the assumed distance of 59 kpc, the 
$3\sigma$ upper limit is 
$\mysim 9\times 10^{33}\; \mathrm{erg}\; \mathrm{s}^{-1}$.

\medskip
\begin{minipage}[l]{0.46\textwidth}
  \plotonefiddle{6cm_chandra.ps}{270}
  \medskip
  \noindent\begin{minipage}{0.99\linewidth} 
    \vskip \abovecaptionskip\footnotesize\noindent
   Fig. 5. ---
    {\it Chandra\/} X-ray data with ATCA 6 cm radio contours.
    The X-ray gray scale is the same as in
    Fig.~\figone.
    The radio contours
    are 0.25,
    0.5, 0.75, 1, 1.5, 2, 2.5, and 3 mJy beam$^{-1}$; the rms noise is
    $\mysim 7.5\times 10^{-2} \;\mathrm{mJy}\;\mathrm{beam}^{-1}$
    \protect\citep{ab93}.
   \par\vskip \belowcaptionskip
  \end{minipage}
\end{minipage}

\section{Comparison with Optical Data}
   \label{sec:comparison.with.optical.data}

\newcommand{\figsix}{6}

We examined {\it Hubble Space Telescope\/} ({\it HST\/}) archival \oiii\ 
images of the remnant.
Figure~\figsix\ 
is a composite color image of the
remnant with the 6 cm radio data in red, the {\it HST\/} {\oiii} data in green, 
and the {\it Chandra\/} X-ray data in light blue.  The {\it Chandra\/} absolute 
coordinates are currently uncertain at a level of at least $\mysim 1 \arcsec$ 
(see \S\ref{sec:x.ray.data}).

%
\medskip
\begin{minipage}[l]{0.50\textwidth}
  \plotone{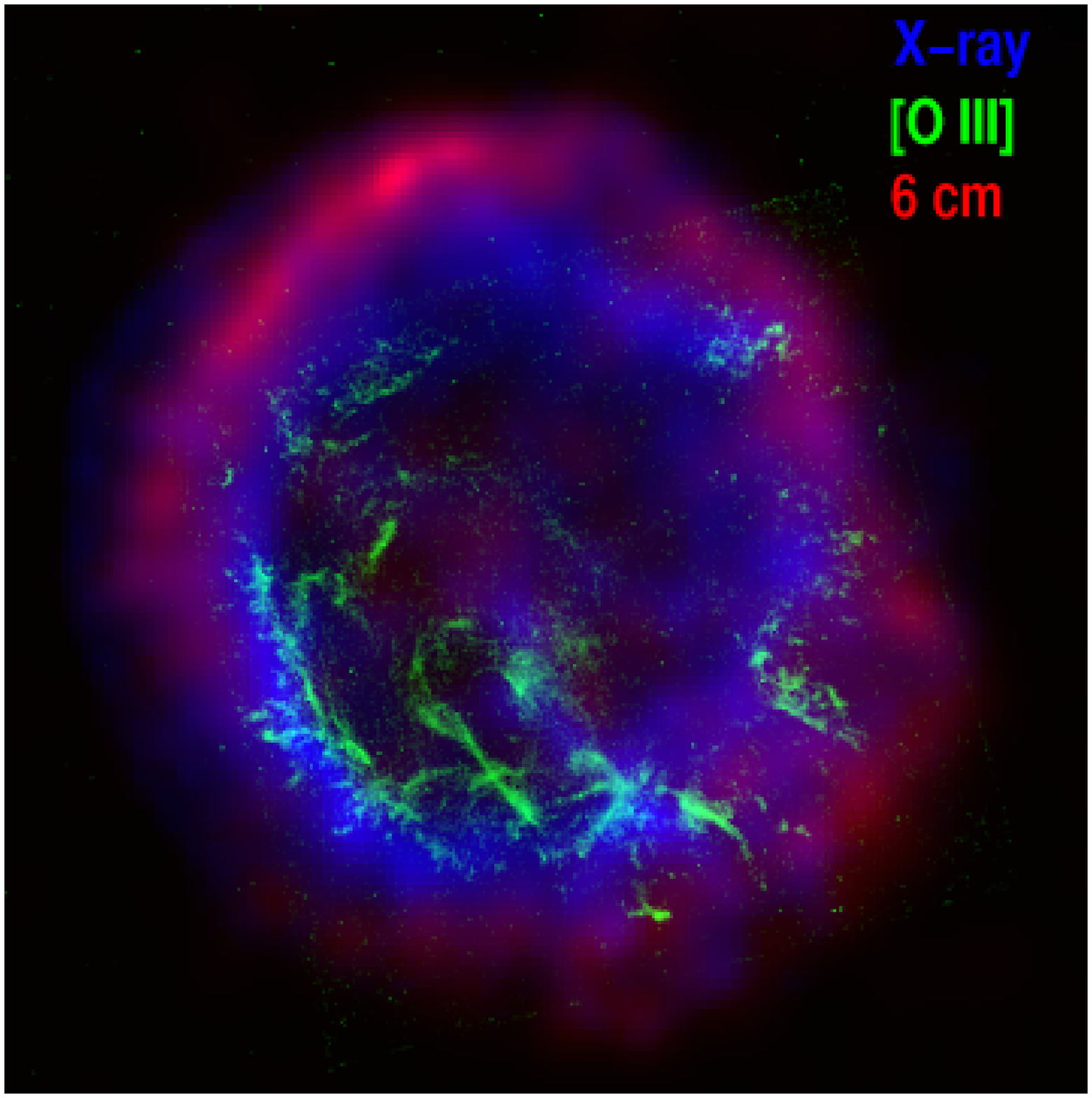}
  \medskip
  \noindent\begin{minipage}{0.99\linewidth} 
    \vskip \abovecaptionskip\footnotesize\noindent
   Fig. 6. ---
     Color composite image of the ATCA 6 cm radio data (red),
     {\it HST\/} {\oiii} data (green), and {\it Chandra\/}
     X-ray data (light blue).
   \par\vskip \belowcaptionskip
  \end{minipage}
\end{minipage}
%

The southeast portion bright X-ray ring coincides with a \break
series of \oiii\  filaments, sharply bounded toward the
center of the remnant, with knots and filaments extending outward.
This suggests that we are seeing a tangency of a shock propagating
through ejecta.  The \oiii\ knots could be engulfed 
dense ejecta clumps or Rayleigh-Taylor fingers of ejecta; the 
optical emission could be produced by shocks driven into cooler,
denser material by the high pressure downstream
of the reverse shock, or by photoionization and excitation by the 
X-rays from the
surrounding material \citep{dbt84, brdm89}.  Note 
that the \oiii\ data were 
obtained with the Planetary Camera
of the {\it HST\/} using the F502N filter.  This filter
is narrow enough that significant emission may be Doppler-shifted 
out of the filter band pass, partic\-ularly for filaments projected 
toward the interior on the front and back faces of an expanding 
shell (Blair et al. 2000).
The inner edges of the southeast \oiii\ filaments may lie inward 
of the X-ray ring by $\mysim 1\arcsec$ (Fig.~\figfour);
however, elsewhere in the remnant, bright \oiii\ knots
coincident with bright X-ray features also tend to be slightly
westward of the X-ray emission.  This may also be in part a result of 
the {\it Chandra\/} absolute coordinate uncertainty.

A number of bright X-ray knots west through south coincide
with optically bright features; most notably, the bright 
X-ray knot in the south-southwest is likely related to the bright \break \oiii\
filament,
the brightest \oiii\ feature in the remnant.  \break 
Bright \oiii\ filaments
border a remarkable gap in the southwest that is suggestive of a
blowout.  The X-ray ring is partic\-ularly broad at this point, and the
radio shows an enhancement as well.  As noted above, the outer
shock in the southwest \break
quadrant is more ragged and less well defined.

The brightest part of the  X-ray ridge in the northwest also coincides
with enhanced \oiii\ emission.  It is also notable that
the 6 cm radio emission shows a ridge in the same location, cupped around
the X-ray and optically bright emission;
the morphology suggests a denser clump of ejecta being 
engulfed by the reverse shock.

There are also puzzling differences between the optical \oiii\ 
and X-ray emission; for example, an optical filament in the southeast
(between the central spoke and the ring) is not associated with
enhanced X-ray emission.  It is also notable that the X-ray and radio
emission look like complete (if nonuniform) rings approximately
in the plane of the sky, while the optical emission shows,
at best, an incomplete ring with large gaps ({\it e.g.\/}, north, west, 
southwest);
\citet{td83} modeled the \oii\ emission and velocity distribution
as an extremely distorted
ring perpendicular to the plane of the sky.
\strut{}
\vskip0.8in

\section{Discussion}
   \label{sec:discussion}

It is plausible to identify the sharp outer edge of the X-ray emission
as the location of the direct blast-wave interacting with the ambient 
medium and the inner edge of the bright ring as the reverse shock.  
The scalloping in the region between the bright ring and the rim might 
then be the result of Rayleigh-Taylor instabilities of the contact surface.
The outer shock is brightest in the northeast, and at least 5  
times fainter in the southwest.  The nonthermal radio emission would be
synchrotron emission originating behind the main blast-wave.
The bright radio ridge in the northeast, corresponding to the brightest 
section of the outer X-ray shock, together with the slight elongation 
northeast to southwest elongation of the remnant suggests that 
the ambient medium may be slightly stratified in that direction.

In the X-ray difference maps (e.g., 
Fig.~\figthree),
the remnant shows a clear layered morphology, 
in which the inner edge of the bright ring is enhanced in \ovii\ emission, 
while the outer part is enhanced in \oviii.  The \neix\ and \nex\ 
emission shows a similar pattern with \neix\ more prominent near the 
inner edge.  The Ne emission is relatively more prominent outward 
of the peak in the \ion{O}{7}-dominated band.  This pattern suggests 
that we are seeing the ionization of the gas which has passed through 
the reverse shock.  The southeast portion of the ring is particularly 
interesting in
this regard, being much brighter and narrower than the rest of the ring;  
Figure \figfour\ 
shows radial profiles
through this portion of the remnant.  The steep
rise at the inner edge and the lag of \oviii\ relative to \ovii\ are 
suggestive of a reverse shock, possibly propagating through a steep
density gradient. 

In summary, the {\it Chandra\/} observations have allowed us to detect 
the outer blast wave for the first time, and to determine accurately the size 
of the remnant; the ring of radio emission
is predominantly inward of the X-ray blast wave, which is consistent with 
an interpretation as synchrotron radiation originating behind the
blast wave, but outward of the bright X-ray ring of emission.  
Complex variations are seen in the bright X-ray ring.
The X-ray spectra vary throughout the remnant.  Difference images
centered on the prominent H-like and He-like lines of O and Ne
suggest that the bright ring corresponds to an ionizing 
reverse shock.  Many (but not all) of the
prominent optical filaments are seen to correspond to X-ray bright
regions.



\acknowledgments

We thank Shaun Amy and Lewis Ball for making available their published 
6 cm data.  We also thank the referee, Jon Morse, for suggestions
which markedly improved this article.  This work was supported by 
NASA contract NAS8-39073, and in part by NASA LTSA NAG5-3559.


\end{document}